%% file: gks.tex
\documentclass[12pt]{amsart}
\usepackage{amsmath,amssymb,amsthm}
\usepackage{epsfig,graphicx,verbatim}
\usepackage{color}

\usepackage{url,verbatim}
\RequirePackage[colorlinks,citecolor=blue,urlcolor=blue]{hyperref}
\usepackage{breakurl}

\numberwithin{equation}{section}
\newtheorem{thm}[equation]{Theorem}

\newcounter{mycount}

\newcommand{\uinvnorm}{|\kern-2pt|\kern-2pt|}

\def\1{\mbox{1\hskip-.25em l}}
\newcommand{\beq}{\begin{equation}}
\newcommand{\eeq}{\end{equation}}

\theoremstyle{plain}

\theoremstyle{definition}

\theoremstyle{remark}

\def\sF{\mathcal F}

\def\sQ{\mathcal Q}

\def\qq{\qquad}

\def\de{\delta}

\def\s{\sigma}

\def\rc{random cluster}
\def\EE{\mathbb E}
\def\la{\langle}
\def\ra{\rangle}

\def\Si{\Sigma}

\def\RR{{\mathbb R}}

\def\CC{{\mathbb C}}
\def\Om{\Omega}

\def\om{\omega}

\def\oo{\infty}

\def\bigmid{\,\big|\,}
\def\fq{\sF_q}
\def\nequiv{\not\equiv}

\def\be{\begin{equation}}
\def\ee{\end{equation}}
\def\sm{\setminus}

\def\lra{\leftrightarrow}
\def\nlra{\nleftrightarrow}

\newcommand\ZRC{Z_{\mathrm{RC}}}

\begin{document}

\title[Correlation inequalities for the Potts model]
{Correlation inequalities\\ for the Potts model}
\author{Geoffrey R. Grimmett}
\address{Statistical Laboratory,
Centre for Mathematical Sciences,
University of Cambridge,
Wilberforce Road, Cambridge CB3 0WB, U.K.}
\email{\url{g.r.grimmett@statslab.cam.ac.uk}}
\urladdr{\url{http://www.statslab.cam.ac.uk/~grg/}}
\date{2 December 2015} %%\today}

\dedicatory{Dedicated in friendship to Lucio Russo}

\begin{abstract}
Correlation inequalities are presented for ferromagnetic Potts models 
with external field, using the \rc\ representation
of Fortuin and Kasteleyn, together with the FKG inequality.
These results extend and simplify earlier inequalities
of Ganikhodjaev and Razak, and also of  Schonmann, and include
GKS-type inequalities when the spin-space
is taken as the set of $q$th roots of unity.
\end{abstract}
\maketitle

\section{Introduction}\label{sec:int}

Correlation inequalities are key to the classical theory of interacting systems in statistical mechanics. 
The Ising model, especially, has a plethora of associated inequalities which have
played significant roles in the development of a coherent theory of phase transition
(see, for example, the books \cite{FFS,MW73}). These inequalities are frequently named after
their discoverers, and include inequalities of Griffiths \cite{Griff1,Griff2},
Griffiths, Kelly, and Sherman (GKS) \cite{KS}, Griffiths, Hurst, and Sherman (GHS) \cite{GHS},
Ginibre \cite{Gin70}, Simon and Lieb \cite{Lieb80,Simon}, and so on.

A more probabilistic theory of Ising/Potts
models has emerged since around 1970, initiated partly by the work of Fortuin and Kasteleyn 
\cite{F72,F72b,FK72} on the 
\emph{random cluster} representation of the Potts model, and the 
\emph{random current method} championed
by Aizenman \cite{Aiz82} and co-authors.
Probably the principle inequality in the probabilistic formulation is
that of Fortuin, Kasteleyn, and Ginibre (FKG) \cite{FKG}.

Inequalities are rarer for the Potts model, and our purpose in this 
note is to derive certain correlation inequalities 
for a ferromagnetic Potts model with external field, akin to the GKS inequalities
for the Ising model. The main technique used here
is the \rc\ representation of this model, and particularly the FKG inequality.

Our results generalize and simplify the work of Ganikhodjaev and Razak
\cite{GanR}, who have shown how to formulate and prove
GKS-type inequalities for the Potts model with a general number $q$
of local states. Furthermore, our Theorems \ref{mainthm} and \ref{s2}
extend the two correlation inequalities of Schonmann \cite{S88}, which in turn
extended inequalities of \cite{DMMR}. 
Some of the arguments given here may be known to others. 

The structure of this paper is as follows. 
The Potts and \rc\ models are introduced in Section \ref{sec:ineq},
and the results of the paper (Theorems \ref{mainthm}--\ref{s2}) follow in Section \ref{sec:ineq2}.
The proofs are given in Sections \ref{sec:pf}--\ref{s2pf}.

\section{The Potts model with external field}\label{sec:ineq}

Let $G=(V,E)$ be a finite graph, and let $J=(J_e: e\in E)$ and $h=(h_v: v \in V)$ be 
vectors of non-negative reals,
and $q \in \{2,3,\dots\}$. An edge $e\in E$ joins two distinct vertices $x$, $y$, and we write
$e=\la x,y\ra$.

We take as `local state space' for the $q$-state Potts model the set
$\sQ:=\{0,1,\dots,q-1\}$ of `spins'. The configuration space of the model is 
the product space $\Si:=\sQ^V$, and a typical configuration is written $\s=(\s_v: v \in V)\in \Si$. 
The  Potts measure on $G$
with parameters $J$, $h$ has sample space $\Si$ and probability measure
given by
$$ 
\pi(\s) = \frac1Z \exp\left\{ \sum_{e=\la x,y\ra\in E} J_e \de_e(\s) + \sum_{v\in V} h_v \de_v(\s)\right\},
\qq \s\in\Si,
$$
where 
$\de_e(\s) = \de_{\s_x,\s_y}$
and $\de_v(\s) = \de_{\s_v,0}$ are Kronecker delta functions, and $Z$ is the 
appropriate normalizing constant. Thus, the $J_e$ are \emph{edge coupling constants},
and the $h_v$ are \emph{external fields} relative to the local state $0$.
The Potts measure is said to be \emph{ferromagnetic} since $J_e\ge 0$ for $e \in E$.

We shall make use of the \rc\ representation, of which we refer
the reader to \cite{G-RC} for a recent account and bibliography.
The graph $G$ is augmented by adding a `ghost' vertex $g$, 
which is joined by edges $\la g,v\ra$ to
each vertex $v \in V$; the ensuing graph is denoted $G^+=(V^+,E^+)$. 
The relevant sample space is the product space $\Om:=\{0,1\}^{E^+}$.
For $\om=(\om_e:e\in E^+)\in\Om$, an edge $e$ is called \emph{open}
if $\om_e=1$ and \emph{closed} otherwise. 

An edge $e \in E$ is assigned parameter $p_e=1-e^{-J_e}$,
and an edge of the form $\la g,v\ra$  is assigned parameter $p_v=1-e^{-h_v}$. 
The \rc\ probability measure $\phi$ on $G$ has sample space $\Om$ and is given by
$$
\phi(\om)= \frac1{\ZRC}\left\{\prod_{e=\la x,y\ra \in E^+} p_e^{\om_e}(1-p_e)^{1-\om_e}\right\}
q^{k(\om)},\qq \om\in\Om,
$$
where $k(\om)$ is the number of connected components
of the graph with vertex set $V^+$ and edge set $\eta(\om):=\{e\in E^+: \om_e=1\}$.

The relationship between the Potts model and the \rc\ model is explained in
\cite[Sect.~1.4]{G-RC}, where it is shown in particular that $\ZRC=e^{-|E|}Z$.

The measures $\pi$ and $\phi$ may be coupled as follows.
Suppose $\om$ is sampled from $\Om$ according to $\phi$, and let $C_v$ be the 
connected component of
$(V,\eta(\om))$ containing $v \in V^+$; the $C_v$ are called \emph{open clusters}.
Every vertex in $C_g$ is allocated spin $0$. To an open cluster of $\om$
other than $C_g$, we allocate a uniformly chosen spin from $\sQ$, such that every vertex
in the cluster receives this spin, and the spins of different clusters
are independent. The ensuing spin vector $\s=\s(\om)$ has law $\pi$.
See \cite[Thm 1.3]{G-RC} for a proof of this standard fact, and for 
references to the original work of Fortuin and Kasteleyn.

Use will be made in this paper of the FKG inequality and the comparison inequalities for the \rc\ model.
These are presented in a number of places already, and are not repeated here. The reader is referred
instead to \cite[Thm 3.8]{G-RC} for the FKG inequality, and to \cite[Thm 3.21]{G-RC} for the comparison
inequalities.

\section{The correlation inequalities}\label{sec:ineq2}

We begin with a space of functions.
Let $\fq$ be the set of  functions $f:\sQ\to\CC$ such that,
for all integers $m,n\ge 0$,
\begin{gather}
\EE(f(X)^m) \text{ is real and non-negative},\label{2}\\
\EE(f(X)^{m+n}) \ge \EE(f(X)^m) \EE(f(X)^n),\label{3}
\end{gather}
where $X$ is a uniformly distributed random variable on $\sQ$.
The above conditions may be written out as follows.
We have that $f \in \fq$ if, for $m,n\ge 0$,
\begin{align*}
&S_m:=\sum_{x\in\sQ}f(x)^m \text{ is real and non-negative}, \\
&qS_{m+n} \ge S_m S_n.
\end{align*}

For $I\in\sQ$, let $\fq^I$ be the subset of $\fq$ containing all $f$ such that
\begin{equation}
f(I)=\max\{|f(x)|: x \in \sQ\}.
\label{1}
\end{equation}
This condition entails that $f(I)$ is real and non-negative.

Let $f:\sQ\to\CC$.
For $\s\in\Si$, let 
\be
f(\s)^R:=\prod_{v\in R} f(\s_v),\qq R\subseteq V.
\label{sigprod}
\ee
Thinking of $\s$ as a random vector with law $\pi$, we write
$\la f(\s)^R\ra$ for the mean value of $f(\s)^R$.

\begin{thm}\label{mainthm}
Let $f\in\fq^0$.
For $R\subseteq V$, the mean $\la f(\s)^R\ra$  is real-valued and non-decreasing in the vectors
$J$ and $h$, and satisfies $\la f(\s)^R\ra \ge 0$. For $R,S\subseteq V$, we have that 
$$
\la f(\s)^R f(\s)^S\ra \ge \la f(\s)^R\ra \la f(\s)^S\ra.
$$
If there is no external field, in that $h\equiv 0$, it suffices for the above
that $f \in \fq$ in place of $f\in \sF_q^0$.
\end{thm}

Here are three classes of functions belonging to $\sF_q^0$.

\begin{thm}\label{pi}
 Let $q \ge 2$.
The following functions $f:\sQ\to\CC$ belong to $\fq^0$.
\begin{itemize}
\item[(a)]  $f(x) = \frac12(q-1)-x$.
\item[(b)] $f(x) = e^{2\pi ix/q}$, a $q$th root of unity.
\item[(c)] $f:\sQ\to[0,\oo)$, with $f(x)\le f(0)$ for  $x\in\sQ$.
\end{itemize}
\end{thm}

When combined with Theorem \ref{mainthm}, 
case (a) yields the inequalities of Ganikhodjaev and Razak \cite{GanR}, but with simpler proofs. 
When $q=2$, the latter reduce
to the GKS inequalities for the Ising model, 
see \cite{Griff1,Griff2,KS}. We do not know
if the implications of Theorem \ref{mainthm} with case (b) are either known or useful.
Perhaps they are examples
of the results of Ginibre \cite{Gin70}.  In case (c) 
with $f(x)=\de_{x,0}$, Theorem \ref{mainthm} yields the first
correlation inequality of Schonmann \cite{S88}.

Our second main result follows next.

\begin{thm}\label{s2}
Let $q\ge 2$, $f_0\in\fq^0$, and let $f_1:\sQ\to\CC$ satisfy \eqref{2}.
If $f_0$ and $f_1$ have disjoint support
in that $f_0f_1\equiv 0$ then, for $R,S\subseteq V$,
$$
\la f_0(\s)^{R}f_1(\s)^{S}\ra \le \la f_0(\s)^{R}\ra \la f_1(\s)^{S}\ra.
$$
If $h \equiv 0$, it is enough to assume $f_0\in\fq$ in place of $f_0\in\sF_q^0$.
\end{thm}

Two correlation inequalities were proved in \cite{S88}, a `positive'
inequality that is implied by Theorems \ref{mainthm} and \ref{pi}(c), and a `negative'
inequality that is obtained as a special case of Theorem \ref{s2},
on setting $f_0(x)=\de_{x,0}$ and $f_1(x) = \de_{x,1}$.
Recall that Schonmann's inequalities were themselves (partial) generalizations
of correlation inequalities of \cite{DMMR}.

Amongst the feasible extensions of the above theorems that come to mind, we mention
the classical space--time models used to study the quantum Ising/Potts models, 
see \cite{aizenman_nacht,JBj2,BjG,CrI,grimmett_stp}.

\section{Proof of Theorem \ref{mainthm}}\label{sec:pf}

We use the coupling of the \rc\ and Potts model described in Section \ref{sec:ineq}.
Let 
$\om\in\Om$, and let $A_g,A_1,A_2,\dots,A_k$ be the vertex-sets
of the open clusters of $\om$, where $A_g$ is that of
the open cluster $C_g$ containing $g$. 

Let $R \subseteq V$, and let $f \in \fq^0$. By \eqref{sigprod}, 
$$
f(\s)^R = f(0)^{|R\cap A_g|}\prod_{r=1}^{k} f(X_r)^{|R\cap A_r|},
$$
where $X_r$ is the random spin assigned to $A_r$. This has
conditional expectation $g_R:\Om\to\CC$ given by
\begin{align*}
g_R(\om) &:= \EE\bigl(f(\s)^R\bigmid\om\bigr)\\ 
&= f(0)^{|R\cap A_g|}\prod_{r=1}^k \EE\bigl(f(X)^{|R\cap A_r|}\bigmid\om\bigr).
\end{align*}
By \eqref{2} and \eqref{1}, $g_R(\om)$ is real and non-negative, whence
so is its mean $\phi(g_R) = \la f(\s)^R\ra$. (It will be convenient to use $\phi(Y)$ to denote 
the expectation of a random variable $Y:\Om\to\RR$.)

We show next that $g_R$ is a non-decreasing function on the partially ordered set $\Om$.
It suffices to consider the case when 
the configuration $\om'$ is
obtained from $\om$ by adding an edge between two clusters of $\om$. In this case,
by \eqref{3}--\eqref{1}, $g_R(\om') \ge g_R(\om)$. 
That $\la f(\s)^R\ra = \phi(g_R)$ is non-decreasing in $J$ and $h$ follows
by the appropriate comparison inequality for the \rc\ measure $\phi$,
see \cite[Thm 3.21]{G-RC}. 

Now,
\begin{equation*}
\EE\bigl(f(\s)^Rf(\s)^S\bigmid \om\bigr) =f(0)^{|R\cap A_g| + |S\cap A_g|}
\prod_{r=1}^k \EE\bigl(f(X)^{|R\cap A_r|+|S\cap A_r|}\bigmid\om\bigr).
\end{equation*}
By \eqref{3},
$$
\EE\bigl(f(\s)^Rf(\s)^S\bigmid \om\bigr) \ge g_R(\om) g_S(\om).
$$
By the FKG property of $\phi$, see \cite[Thm 3.8]{G-RC},
\begin{align*}
\la f(\s)^Rf(\s)^S\ra &= \phi\bigl(\EE\bigl(f(\s)^Rf(\s)^S\bigmid\om\bigr)\bigr)\\
&\ge \la f(\s)^R\ra \la f(\s)^S\ra,
\end{align*}
as required.

When $h \equiv 0$, the terms in $f(0)$ do not appear in the above,
and it therefore suffices that $f\in\fq$.

\section{Proof of Theorem \ref{pi}}

We shall use the following elementary fact: if $T$ is a non-negative
random variable, 
\be
\EE(T^{m+n}) \ge \EE(T^m)\EE(T^n),\qq m,n\ge 0.
\label{triv}
\ee
This trivial inequality may be proved in several ways, of which
one is the following. Let $T_1$, $T_2$ be independent copies of $T$.
Clearly,
\begin{equation}\label{eq:3}
(T_1^m-T_2^m)(T_1^n-T_2^n) \ge 0,
\end{equation}
since either $0\le T_1\le T_2$ or $0\le T_2 \le T_1$. Inequality \eqref{triv}
follows by multiplying out \eqref{eq:3} and averaging.

\medskip
\noindent
\emph{Case }(a).
Inequality \eqref{1} with $I=0$ is a triviality.
Since $f(X)$ is real-valued, with the same distribution as $-f(X)$, $\EE(f(X)^m) =0$
when $m$ is odd, and is positive when $m$ is even.
When
$m+n$ is even, \eqref{3} follows from \eqref{triv} with $T=f(X)^2$, 
and both sides of \eqref{3} are $0$ otherwise.

\medskip
\noindent
\emph{Case }(b).
It is an easy calculation that
$$
\EE(f(X)^m) = \begin{cases} 1 &\text{if } q \mid m,\\
0 &\text{otherwise},
\end{cases}
$$
and \eqref{2}--\eqref{3} follow.

\medskip
\noindent
\emph{Case }(c). Inequality \eqref{3} follows by \eqref{triv} with
$T=f(X)$.

\section{Proof of Theorem \ref{s2}}\label{s2pf}

We may as well assume that $f_0\nequiv 0$, so that $f_0(0)>0$
and $f_1(0)=0$.
We use the notation of Section \ref{sec:pf}, and let $F_i:\Om\to\CC$ be given by
\begin{align}
F_0(\om) &= f_0(0)^{|R\cap A_{g}|} \prod_{r=1}^k 
\EE\bigl(f_0(X)^{|R\cap A_r|}\bigmid\om\bigr),\label{mel4}\\
F_1(\om) &=  \prod_{r=1}^k \EE\bigl(f_1(X)^{|S\cap A_r|}\bigmid\om\bigr).
\label{mel5}
\end{align}
By \eqref{2}, $F_0$ and $F_1$ are real-valued and non-negative.
Since $f_0\in\fq^0$, $F_0$ is non-decreasing (as in Section \ref{sec:pf}).

Since $f_0f_1\equiv 0$, 
\begin{align*}
\EE\bigl(f_0(\s)^R f_1(\s)^S\bigmid\om\bigr) 
= 1_Z(\om) F_0(\om)F_1(\om),
\end{align*}
where $1_Z$ is the indicator function of the event 
$Z= \{S \nlra R\cup\{g\}\}$.
Here, as usual, we write $A\lra B$ if there exists an open path in $\om$ from some vertex
of $A$ to some vertex of $B$.
Let $T$ be the subset of $V^+$ containing all vertices joined to $S$ by open paths,
and write $\om_T$ for the configuration $\om$ restricted
to $T$. Using conditional expectation,
\begin{align}
\la f_0(\s)^Rf_1(\s)^S\ra &= \phi\bigl( 1_Z F_0 F_1\bigr)\label{m1}\\
&=\phi\bigl( 1_Z F_1 \phi(F_0\mid T,\,\om_T)\bigr),
\nonumber
\end{align}
where we have used the fact that $1_Z$ and $F_1$ 
are functions of the pair $T$, $\om_T$ only.
On the event $Z$, 
$F_0$ is a non-decreasing function of the configuration restricted to $V^+ \sm T$.
Furthermore, given $T$, the conditional measure on $V^+ \sm T$ is
the corresponding \rc\ measure.  It follows that
$$
\phi(F_0\mid T,\,\om_T) \le \phi(F_0)\quad \text{on the event } Z,
$$
by \cite[Thm 3.21]{G-RC}.
By \eqref{m1},
\begin{align*}
\la f_0(\s)^Rf_1(\s)^S\ra &\le \phi\bigl( 1_Z F_1 \phi(F_0)\bigr)\\
&\le \phi(F_0)\phi(F_1)\\
&= \la f_0(\s)^R\ra\la f_1(\s)^S\ra ,
\end{align*}
and the theorem is proved.

When $h\equiv 0$, $A_g = \{g\}$ in \eqref{mel4}, and it suffices that $f_0 \in \fq$.

\section*{Acknowledgements}
This work was supported in part
by the Engineering and Physical Sciences Research Council under grant EP/103372X/1. 
The author is grateful to Jakob Bj\"ornberg for proposing the model
with angular spins, and to Chuck Newman, and Aernout van Enter also
for their comments and suggestions. 

\input{gks.bbl}

%\bibliography{griffiths2}
\bibliographystyle{amsplain}

\end{document}

%% file: gks.bbl
\providecommand{\bysame}{\leavevmode\hbox to3em{\hrulefill}\thinspace}
\providecommand{\MR}{\relax\ifhmode\unskip\space\fi MR }
\providecommand{\MRhref}[2]{%
  \href{http://www.ams.org/mathscinet-getitem?mr=#1}{#2}
}
\providecommand{\href}[2]{#2}